\documentstyle[aps]{revtex}
\begin{document}
\draft

\title{Quantum-Classical Correspondence  
for the Equilibrium Distributions 
of Two Interacting Spins}
\author{J. Emerson and L.E. Ballentine}
\address{Physics Department, Simon Fraser University, 
Burnaby, British Columbia, Canada V5A 1S6}
\date{\today}
\maketitle
\begin{abstract}
We consider the quantum and classical Liouville dynamics 
of a non-integrable model of two coupled spins.  
Initially localised quantum states  
spread exponentially to the system dimension when the classical 
dynamics are chaotic. 
The long-time behaviour of the quantum probability distributions
and, in particular, the parameter-dependent rates of relaxation to the 
equilibrium state are surprisingly well approximated by the 
classical Liouville mechanics even for small quantum numbers.
%If the classical dynamics are chaotic, the quantum 
%steady-states ((equilibrium states))  
%exhibit rapidly oscillating fluctuations away from 
%the corresponding classical steady-state ((equilibrium state)). 
As the accessible classical phase space becomes predominantly chaotic, 
the classical and quantum probability equilibrium configurations approach 
the microcanonical distribution, although the  
quantum equilibrium distributions exhibit characteristic 
`minimum' fluctuations 
%((or `characteristic' or `irreducible')) standard deviation 
away from the microcanonical state.    
The magnitudes of the quantum-classical differences 
arising from the equilibrium quantum fluctuations 
are studied for both pure and mixed (dynamically entangled) quantum states. 
In both cases the standard deviation of these fluctuations 
decreases as $(\hbar/{\mathcal J})^{1/2}$, where ${\mathcal J}$ 
is a measure of the system size.
%Even for the pure states of this 
In conclusion, under a variety of conditions 
the differences between quantum and classical Liouville mechanics 
are shown to 
become vanishingly small in the classical limit 
(${\mathcal J}/\hbar \rightarrow \infty$) 
of a non-dissipative model endowed with only a few degrees of freedom.
% model that is endowed with only a few classical 
%degrees of freedom. 
\end{abstract}

\section{Introduction}

The study of chaos in quantum dynamics has led to differing views 
on the conditions required for demonstrating 
quantum-classical correspondence \cite{Ball94,Zurek98a}.   
Moreover, the criteria by which this correspondence should be 
measured have also been a subject of some 
controversy \cite{ZP94,CC95,ZP95a}.    
%and the emergence of classical behaviour 
%especially in the macroscopic limit 
While much of the earlier work 
on this topic is concerned with characterizing 
the degree of correspondence between quantum expectation values and 
classical dynamical variables \cite{BZ78,Frahm85,Haake87}, the 
more recent approach is to focus on differences between 
the properties of quantum states and associated 
classical phase space densities 
evolved according to Liouville's equation 
\cite{Ball94,HB94,Fox94b,RBWG95,Ball98,EB00a}.  

Several authors have examined quantum-classical correspondence 
by considering the effects of interactions with a stochastic 
environment \cite{Dec96,HB96,Brumer99}, a process sometimes 
called {\em decoherence}. While this process may improve the degree 
of quantum-classical correspondence for fixed quantum numbers, 
it has been further suggested that the  
limit of large quantum numbers is inadequate for correspondence, 
and that decoherence {\em must} be taken into account 
to generate classical appearances from quantum theory; 
this view has been argued to apply  
even in the case of macroscopic bodies that are described 
initially by well-localised states, 
provided their classical motion is chaotic \cite{ZP95a,ZP95b,Zurek98b}. 
In this paper we examine how the degree of correspondence 
with Liouville dynamics scales specifically 
in the limit of {\em large} quantum numbers. 
This ``classical limit'' is distinct from a ``thermodynamic 
limit'', that is, a limit involving {\em many} quantum numbers. 

The degree of Liouville correspondence has been 
characterized previously by studying 
the differences between the means and variances of the 
dynamical variables \cite{Ball94,HB94,RBWG95,Ball98,EB00a,Ball00}. 
This involves a comparison of quantum expectation values and classical
ensemble averages. 
However, these low-order moments give only crude information about the 
differences between the quantum and classical states. 
Specifically, the quantum state may exhibit coarse structure which 
differs significantly 
from the classical state although the  
means and variances (for some simple observabes) 
are nearly the same for the quantum and classical states. 
Moreover, much of the previous work was concerned with correspondence 
at early times, or more precisely, in the Ehrenfest regime when the 
states are narrow compared to system dimensions \cite{Ball98,EB00a}.

Another approach is to identify quantum-classical differences with 
differences between the Wigner quasi-distribution and the 
classical phase space density \cite{Zurek98a}. 
This approach is objectionable 
because the Wigner quasi-distribution may take 
on negative values and therefore may not be interpreted as 
a ``classically observable'' phase space distribution. 
It is possible to consider instead smoothed quantum phase space 
distributions, but in this case the residual quantum-classical  
differences still do not have clear experimental  
significance.  

In this paper, we characterize the degree of 
quantum-classical correspondence by comparing quantum 
probability distributions for dynamical variables 
with the corresponding classical marginal 
distributions for these dynamical variables. 
These are well-defined classical observables that describe the 
distribution of outcomes upon measurement of the 
given dynamical variable.  We are interested in the 
differences that arise on a {\em fine} scale and therefore 
characterize the typical quantum-classical deviations  
that arise in bins of width $\hbar$.  

The dynamics are generated by a model of interacting spins that  
we have studied previously in Ref.\ \cite{EB00a}.  The Hilbert space
is finite dimensional so no artificial truncation of the state is required. 
The quantum time-evolution is unitary and the classical 
motion is volume-preserving (symplectic).
In the case of classically chaotic motion, we follow 
initially localised states until they have evolved well beyond 
the relaxation time-scale of the classical density. 
Throughout the paper we emphasize that the {\em quantum} signatures of 
chaos that appear in the quantum distributions 
are the same as those that appear in 
the marginal classical distributions. 
In particular, the quantum relaxation rates can be accurately estimated 
from the Liouville dynamics of an approximately 
matching initial phase space density. 
This purely classical approximation is surprisingly accurate  
even for small quantum numbers, 
but may be most useful for the theoretical description of 
mesoscopic systems since 
the purely classical calculations do not scale with the quantum numbers.   

The quantum and classical 
probability distributions remain close even after the states 
have spread to the system dimension. 
Specifically, in mixed regimes, 
the quantum distributions exhibit an 
equilibrium shape that reflects the details 
of the classical KAM surfaces.  
When the classical manifold is predominantly chaotic, the quantum  
and classical states relax close to the microcanonical state.  
However, in both of these chaotic regimes the equilibrium quantum 
distributions exhibit characteristic 
fluctuations away from the classical ones.  
We demonstrate that 
the standard deviation of these quantum-classical differences becomes 
vanishingly small in the classical limit, 
${\mathcal J}/\hbar \rightarrow \infty$, where 
${\mathcal J}$ is a characteristic system dimension. 

This paper is organised as follows. 
In section II we describe the quantum and classical 
models for our system. In section III we describe 
the initial quantum states and corresponding classical 
densities. In section IV we give an overview of the 
dynamics of the probability distributions in 
the different classical regimes. In section V we examine the quantum 
and classical relaxation rates using the Shannon entropy. 
%In section VI we examine the degree of deviation 
%from microcanonical equilibrium that arisis in the different 
%classical regimes. 
In section VI we give an overview of the time-development of the small 
differences between the quantum and 
classical probability distributions. In section VII we 
show that the relative quantum-classical differences 
decrease as an inverse power of the Hilbert space dimension.
In section VIII we provide a brief summary and 
explain how our results inform current discussions of 
the necessary and sufficient conditions 
for the emergence of classical properties from quantum mechanics.

\section{The Model}
\label{sec2}

We consider the quantum and classical dynamics 
generated by a non-integrable model of two interacting spins, 
\begin{equation}
        H = a (S_z + L_z) + c S_x L_x \sum_{n = -\infty}^{\infty}
\delta(t-n ) 
\label{eqn:ham}
\end{equation}
where ${\bf S} = (S_x,S_y,S_z)$ and 
${\bf L} = (L_x,L_y,L_z)$.
The first two terms correspond to simple rotation of both spins 
about the $z$-axis through an angle $a$ with range $2\pi$ radians. 
The sum over coupling terms describes an infinite 
sequence of $\delta$-function interactions 
at times $t=n$ for integer $n$. 
Each interaction term produces an impulsive rotation 
of each spin about the $x$-axis by an angle proportional 
to the $x$-component of the other spin. 

\subsection{The Quantum Dynamics}

%We consider the quantum
%dynamics of a non-integrable system of two 
%coupled angular momenta, which we denote ${\bf S}$ 
%and ${\bf L}$. The dynamics are generated by the 
%Hamiltonian,

To obtain the quantum dynamics we interpret 
the Cartesian components of the spins 
as operators satisfying the usual angular momentum 
commutation relations,  
%${\bf S} = S_x {\bf i} + S_y {\bf j}+ S_z {\bf k} $ and 
%${\bf L} = L_x {\bf i} + L_y {\bf j}+ L_z {\bf k} $ 
\begin{eqnarray*}
  & [ S_i,S_j ] = & i \epsilon_{ijk} S_k \\
  & [ L_i,L_j ] = & i \epsilon_{ijk} L_k \\ 
  & [ J_i,J_j ] = & i \epsilon_{ijk} J_k .
\end{eqnarray*}
In the above we have set $\hbar =1$ and introduced the total 
angular momentum vector ${\bf J} = {\bf S} + {\bf L}$.

The Hamiltonian (\ref{eqn:ham}) possesses kinematic 
constants of the motion, $ [ {\bf S}^2,H] = 0$ and $[ {\bf L}^2,H] =0 $. 
Thus the total state vector $|\psi\rangle$ 
can be represented in an invariant Hilbert space ${\mathcal H} = 
{\mathcal H}_s \otimes {\mathcal H}_l $, with   
dimension $ N=(2s+1) \times (2l+1)$, 
that is spanned by the orthonormal vectors 
\begin{equation}
\label{eqn:basis}
|s,l,m_s,m_l \rangle = |s,m_s \rangle \otimes |l,m_l \rangle 
\end{equation}
with $m_s \in \{ s, s-1, \dots, -s \}$ 
and $m_l \in \{ l, l-1, \dots, -l \}$.  
%These are the joint eigenvectors of the four spin operators 
%\begin{eqnarray}
%\label{eqn:basis}
%	{\bf S}^2 |s, l, m_s,m_l \rangle & =& s(s+1) 
%| s,l,m_s,m_l \rangle  \nonumber \\
%	S_z | s, l,m_s,m_l \rangle & =& m_s 
%| s,l,m_s,m_l \rangle  \\
%	{\bf L}^2 | s,l,m_s,m_l \rangle &= & l(l+1) 
%| s,l,m_s,m_l \rangle  \nonumber \\
%	L_z | s,l,m_s,m_l \rangle &= & m_l 
%| s,l,m_s,m_l \rangle  \nonumber .
%\end{eqnarray}

It should be noted that the 
components of the total angular momentum are not conserved  
$[J_i,H] \neq 0$. 
%, as well as its magnitude, $[{\bf J}^2,H] \neq 0$,  are not conserved.
%This is an unphysical peculiarity of the 
%model which arises also in the classical treatment. 
%A purely quantum mechanical phenomena which applies to 
The $z$-component is subject to the selection 
rule, $\Delta J_z = \{ \pm 2,0 \}$, and  
%This selection rule is manifest when the coupling term 
%is expressed as a sum of raising and 
%lowering operators: $ S_x L_x = S_+ L_+ + S_- L_- + S_+ L_- + S_- L_+$.
consequently the full Hilbert space can be decomposed into two 
invariant subspaces.  

The periodic sequence of interactions introduced by the 
$\delta$-function produces a quantum mapping. 
The time-evolution 
for a single iteration, from just before a kick to just before 
the next, is produced by the unitary transformation, 
\begin{equation}
\label{eqn:qmmap}
	| \psi(n+1) \rangle = F \; | \psi(n) \rangle,  
\end{equation}
where $F$ is the single-step Floquet operator, 
\begin{equation}
\label{eqn:floquet}
        F = \exp \left[ - i a (S_z + L_z) \right] 
                \exp \left[ - i c S_x \otimes L_x \right].  
\end{equation}
%Adopting units such that $\tau =1$, 
The quantum dynamics are thus specified by two 
parameters, $a$ and $c$, 
and two quantum numbers, $s$ and $l$.

\subsection{Classical Map} 

For the Hamiltonian (\ref{eqn:ham}) the corresponding 
classical equations of motion 
are obtained by interpreting the angular momentum components 
as dynamical variables satisfying,   
\begin{eqnarray*}
  & \{ S_i,S_j \} = & \epsilon_{ijk} S_k \\
  & \{ L_i,L_j \} = & \epsilon_{ijk} L_k \\ 
  & \{ J_i,J_j \} = & \epsilon_{ijk} J_k ,
\end{eqnarray*}
with $\{ \cdot,\cdot \}$ denoting the Poisson bracket. 
The periodic $\delta$-function in the coupling term 
can be used to reduce the time-evolution to a stroboscopic mapping 
at times $t=n$, for integer $n$, 
\begin{eqnarray}
\label{eqn:map}
        \tilde{S}_x^{n+1} & = & \tilde{S}_x^n \cos( a) - 
        \left[ \tilde{S}_y^n \cos( \gamma r \tilde{L}_x^n ) - 
        \tilde{S}_z^n \sin( \gamma r 
        \tilde{L}_x^n)\right] \sin( a), \nonumber \\ 
        \tilde{S}_y^{n+1} & = & \left[ \tilde{S}_y^n \cos( 
        \gamma r \tilde{L}_x^n) - \tilde{S}_z^n \sin( \gamma r
        \tilde{L}_x^n ) \right] \cos( a) + 
        \tilde{S}_x^n \sin( a), \nonumber \\
        \tilde{S}_z^{n+1} & = &  \tilde{S}_z^n \cos( 
        \gamma r \tilde{L}_x^n) + \tilde{S}_y^n \sin( 
        \gamma r \tilde{L}_x^n), \\
        \tilde{L}_x^{n+1} & = & \tilde{L}_x^n \cos( a) - 
        \left[\tilde{L}_y^n\cos(\gamma \tilde{S}_x^n) - \tilde{L}_z^n 
        \sin(\gamma\tilde{S}_x^n )\right]\sin(a), \nonumber \\ 
        \tilde{L}_y^{n+1} & = & \left[ \tilde{L}_y^n \cos( \gamma 
        \tilde{S}_x^n) - \tilde{L}_z^n \sin( \gamma
        \tilde{S}_x^n) \right] \cos( a) + 
        \tilde{L}_x^n \sin( a), \nonumber \\
        \tilde{L}_z^{n+1} & = &  \tilde{L}_z^n \cos( 
        \gamma \tilde{S}_x^n )+ \tilde{L}_y^n \sin( \gamma 
        \tilde{S}_x^n). \nonumber 
\end{eqnarray}
Here ${\tilde {\bf L}} = {\bf L} / |{\bf L}|$ , 
${\tilde {\bf S}} = {\bf S} / |{\bf S}|$,   
and we have introduced the parameters 
$ \gamma = c |{\bf S}| $ and $ r = | {\bf L}| / | {\bf S}| $. 
The mapping equations (\ref{eqn:map}) describe 
the time-evolution of (\ref{eqn:ham}) 
from just before one kick to just before 
the next. 

Since the magnitudes of both spins are conserved, 
$ \{ {\bf S}^2,H \} = \{ {\bf L}^2,H \} =0$, 
the stroboscopic motion is actually confined to the four-dimensional 
manifold ${\mathcal P} ={\mathcal S}^2 \times {\mathcal S}^2$, 
which corresponds to the surfaces of two spheres. 
This is manifest when the mapping (\ref{eqn:map}) 
is expressed in terms of the four {\it canonical} coordinates 
${\bf x} = (S_z, \phi_s, L_z , \phi_l )$, where 
$\phi_s = \tan (S_y /S_x)$ and $\phi_l = \tan(L_y/L_x) $. 
We will refer to the mapping (\ref{eqn:map}) in canonical form 
using the shorthand notation ${\bf x}^{n+1} = {\bf F}({\bf x}^n)$. 
It is also useful to introduce a complete set of spherical coordinates 
$ \vec{\theta} = (\theta_s,\phi_s,\theta_l,\phi_l) $ where 
$\theta_s = \cos^{-1} (S_z / |{\bf S}|) $ and 
$\theta_l = \cos^{-1} (L_z / |{\bf L}|) $.

As in the quantum model, 
the components of total angular momentum are not constants 
of the motion $\{J_i,H \} \neq 0$. On the other hand, the quantum 
selection rule $\Delta J_z = \{ \pm 2,0 \}$ 
has no classical analogue.  

%The mapping equations are conveniently expressed 
%in terms of the normalized components, 
%\begin{eqnarray}
%	\tilde{L}_x & = & \sin\theta_l \cos\phi_l, 
%%& \tilde{S}_x & = & \sin\theta_s \cos\phi_s 
%\nonumber \\
%	\tilde{L}_y & = & \sin\theta_l \sin\phi_l, 
%%& &=& 
%\nonumber \\
%	\tilde{L}_z & = & \cos\theta_l, 
%%&  &=& 
%\nonumber
%\end{eqnarray}
%and similarly for ${\tilde {\bf S}}$. 

The mapping (\ref{eqn:map}) on the reduced surface  
${\mathcal P}$ enjoys a rather large 
parameter space. The dynamics are  
determined from three independent dimensionless 
parameters ($a$, $\gamma$, and $r \ge 1$), 
where $\gamma= c |{ \bf S }|$ 
is a dimensionless coupling strength 
and $r = |{\bf L}|/ |{ \bf S }| $ corresponds 
to the ratio of the  magnitudes of the two spins. 
The dependence of the classical behaviour 
on these parameters is described in Ref.\ \cite{EB00a}.

\subsection{The Liouville Dynamics}

We are interested in comparing the 
quantum dynamics generated by (\ref{eqn:qmmap}) with 
the corresponding Liouville dynamics of a classical distribution. 
The time-evolution of a Liouville density is generated by the 
partial differential equation,    
\begin{equation}
\label{eqn:liouville}
{ \partial \rho_c({\bf x},t) \over \partial t } = - \{ \rho_c , H \},   
\end{equation}
where $H$ stands for the Hamiltonian (\ref{eqn:ham})  
and ${\bf x} = (S_z, \phi_s,L_z,\phi_l)$. 

%\begin{eqnarray}
%\label{eqn:soln}
%	\rho_c(S_z,\phi_s,L_z,\phi_l,t) & = &  
%\int {\mathrm d} S_z \int {\mathrm d} \phi_s 
%\int {\mathrm d} L_z \int {\mathrm d} \phi_l \; 
%\delta(S_z - S_z(t)) \nonumber \\
%& & \delta(\phi_s - \phi_s(t)) \; 
%\delta(L_z - L_z(t)) \;
%\delta(\phi_l - \phi_l(t)) \; 
%\rho_c(S_z,\phi_s,L_z,\phi_l,0) , 
%\end{eqnarray}

The solution to (\ref{eqn:liouville}) can be expressed in 
the compact form, 
\begin{equation}
\label{eqn:soln}
	\rho_c({\bf x},t) =   
\int_{\mathcal P} {\mathrm d} \mu({\bf y}) \;
\delta({\bf x} - {\bf x}(t,{\bf y})) \; 
\rho_c({\bf y},0), 
\end{equation}
%given phase space ${\mathcal P }$ 
with measure,
\begin{equation}
\label{eqn:measure} 
d \mu({\bf y}) = d \tilde{S}_z d \phi_s d \tilde{L}_z d \phi_l,
\end{equation}
and where  
the time-dependent functions ${\bf x}(t,{\bf y}) \in {\mathcal P}$ are 
solutions to the equations of motion (\ref{eqn:map}) with 
initial conditions ${\bf y} \in {\mathcal P}$. 
This solution expresses the fact that 
Liouville's equation (\ref{eqn:liouville}) describes the dynamics 
of a classical density 
%$\rho_c({\bf x},t)$ 
of points evolving in phase space under the Hamiltonian flow. 
We exploit this fact to numerically solve 
(\ref{eqn:liouville}) by 
randomly generating initial conditions  
consistent with an initial phase space 
distribution $\rho_c({\bf x},0)$ and 
then time-evolving each of these initial conditions 
using the equations of motion (\ref{eqn:map}). 
%which in the case of (\ref{eqn:ham}) 
%consists of the mapping relations 

\subsection{Correspondence Between Quantum and Classical Models}

For a quantum system specified by the four numbers $\{a,c,s,l\}$ 
we determine the corresponding classical parameters $\{a,\gamma,r\}$ 
by first defining the classical magnitudes in terms of the quantum 
magnitudes, 
\begin{eqnarray}
		|{\bf S}| & = & \sqrt{s(s+1)} \nonumber \\
		| {\bf L}| & = & \sqrt{l(l+1)},  
\end{eqnarray}
where the quantities on the left hand side are the lengths of the 
classical spins and those on the right are the quantum 
numbers. If we set the Hamiltonian coefficients $a$ and $c$ 
numerically equal for both models, then the remaining two 
dimensionless classical parameters are determined,  
\begin{eqnarray}
		r & = & 
\sqrt{l(l+1) \over s(s+1)} \nonumber \\  
		\gamma & = & c \sqrt{s(s+1)}.
\end{eqnarray}

We are interested in extrapolating the behaviour of 
the quantum dynamics in the limit $s \rightarrow \infty$ 
and $l \rightarrow \infty$. This is accomplished by 
studying sequences of quantum models with increasing $s$ and $l$ 
chosen so that $r$ and $\gamma$ are held fixed. 
Since $s$ and $l$ 
are restricted to integer (or half-integer) values, the corresponding 
classical $r$ will actually vary slightly for each member of this sequence
(although $\gamma$ can be matched exactly by varying the quantum 
parameter $c$ slightly).  In the limit $s \rightarrow \infty$ 
and $l \rightarrow \infty$ this variation becomes 
increasingly small since $r= \sqrt{l(l+1)/s(s+1)} \rightarrow l/s$. 
%For ease of reference, the classical $r$ corresponding to 
%each member of the 
%sequence of quantum models is identified with its value in this limit. 
We have examined the effect of the small variations in the value of $r$ 
on the classical behaviour and found this variation to have negligible effect.

\section{Initial States}

%\subsection{Initial Quantum States}

We consider {\em initial} quantum states   
which are pure and separable, 
\begin{equation}
        | \psi(0) \rangle = | \psi_s(0) \rangle \otimes |\psi_l(0) \rangle. 
\end{equation}
The initial state of each subsystem is a directed 
angular momentum state,  
\begin{equation} 
\label{eqn:cs}
%        | \psi_j(0) \rangle = 
| \theta,\phi \rangle 
	= R^{(j)}(\theta,\phi) | j,j \rangle,  
\end{equation} 
where $j$ in this section refers to either $l$ or $s$. 
This is a localised state, {\it i.e}.\ one 
of maximum polarization in the 
direction $(\theta,\phi)$, with expectation values of the 
spin components confined to the surface of a 2-sphere, 
\begin{eqnarray}
        \langle \theta, \phi | J_z | \theta, \phi \rangle & = & j \cos
        \theta \nonumber \\
        \langle \theta, 
\phi |J_x \pm i J_y | \theta, \phi \rangle & = &
        j e^{\pm i\phi} \sin \theta.
\end{eqnarray}

%We use the SU(2) coherent states for the initial state 
%of each spin subsystem \cite{cs}:
%\begin{equation} 
%\label{eqn:cs}
%        | \theta,\phi \rangle = ( 1 + \tau \tau^*)^{-j} 
%        \sum_{m=-j}^j \tau^{j-m} C_j^m | j,m \rangle
%\end{equation} 
%where, 
%\begin{eqnarray}
%        \tau & = & \exp (i \phi) \tan ( \theta / 2 ) \noumber \\ 
%        C_j^m & = & \left[ { (2 j)! \over (j+m)!(j-m)! } \right]^{1/2} 
%\end{eqnarray}
%and 

%The coherent state $| \theta,\phi\rangle$ corresponds to a pure 
%state with maximum polarization $m=j$ in the direction 
%$(\theta,\phi)$. This can be inferred from the properties,
%or, equivalently, from the fact that $| \theta,\phi \rangle$ can be 
%obtained by application of the 
%rotation operator $R(\theta,\phi)$ to the 
%state $|j,j\rangle $. 
%It should be noted that the initial 
%state of the total angular momentum 
%operator, ${\bf J}$, is also a coherent-state 
%when $\theta_s=\theta_l$ and $\phi_s=\phi_l$. 

%Coherent states are widely considered to be the most `classical' states, 
%probably because they are the most localised states, though, as 
%demonstrated in \cite{EB00}, 
%this does not appear to be an inadequate criterion 
%of classical-like behaviour in the quantum dynamics. 
%In fact their support 
%in the classical phase space shrinks to a point in the formal 
%limit $\hbar \rightarrow 0 $. 
The  states (\ref{eqn:cs}) are the SU(2) coherent states, 
which, like their counterparts in the Euclidean 
phase space, are minimum uncertainty states \cite{cs}; 
the normalized variance of the quadratic operator,  
\begin{equation}    
\Delta {\bf \tilde J}^2  = { 
 \langle \theta, \phi | {\bf J}^2 | \theta, \phi \rangle 
- \langle \theta, \phi | {\bf J} | \theta, \phi \rangle^2 
\over  j(j+1) }  = {1 \over  (j+1)},
\end{equation}
is minimised for given $j$ and vanishes 
in the limit $j \rightarrow \infty$. 
The coherent states directed along the $z$-axis, 
$| j,j\rangle $  and $| j,-j\rangle $,  
saturate the inequality of the Heisenberg uncertainty relation,  
\begin{equation}
	\langle J_x^2 \rangle 
	\langle J_y^2 \rangle \ge { \langle J_z \rangle^2 \over 4 }, 
\end{equation}
although this inequality is not saturated 
for coherent states polarized in other directions. 
%The coherent state covers an 
%area of roughly $ \Delta \Omega_l \simeq \pi / (l +1) $ 
%on the unit sphere, which of course 
%also shrinks to zero in the limit $l \rightarrow \infty$. 

%While we do not equate `localised' quantum states with the notion of 
%classical-like quantum states, the coherent states are still of 
%considerable interest 
%since they provide a sufficient condition for  the Ehrenfest regime

%\subsection{Initial Classical States} 

We would like to construct a classical Liouville density on the 2-sphere 
with marginal distributions 
that match the quantum probability distributions.  
But we have shown previously that this is impossible for the 
SU(2) coherent states \cite{EB00a}.  
Thus from the outset it is clear that 
any choice of initial classical state will exhibit residual 
discrepancy in matching some of the initial quantum moments. 

We have examined the correspondence properties of several different 
classical distributions. These included the vector model 
distribution described in the Appendix of \cite{EB00a} 
and the Gaussian distribution 
used by Fox and Elston in correspondence studies of the 
kicked top \cite{Fox94b}. We selected the density,  
\begin{equation}
%\begin{eqnarray}
\label{eqn:rho}
        \rho_c (\theta,\phi) \;  \sin \theta d \theta d \phi =  
\; C
\exp \left[ - { 2 \sin^2({\theta\over 2}) ) 
\over \sigma^2}\right] \sin \theta d \theta d\phi 
%& = & \; C \exp \left[ - {(1 -\tilde{J}_z)\over 
% \sigma^2 } \right] \; d \tilde{J}_z  d\phi, \nonumber 
\end{equation}
%\end{eqnarray}
with $ C = \left[ 2 \pi \sigma^2 \left( 1 - \exp( -2 \sigma^{-2})
 \right) \right]^{-1} $, 
instead of those previously considered 
because it is periodic under $2\pi$ rotation. 
The classical density (\ref{eqn:rho}) has a maximum along the $+z$-axis, 
corresponding to the coherent state $|j,j \rangle$. 
An initial state directed along  
$(\theta_o,\phi_o)$ is produced by a rigid body 
rotation of (\ref{eqn:rho}) 
%\begin{equation}
%\label{eqn:rho}
%\rho_c (\theta-\theta_o,\phi - \phi_o )  \; \sin \theta d \theta d \phi =   
%R_c(\theta_o , \phi_o) \; \rho_c (\theta,\phi)  
%\; \sin \theta d \theta d \phi 
%\end{equation}
%where $R_c(\theta_o,\phi_o)$ corresponds to a 
%rigid body rotation of the distribution $\rho_c(\theta,\phi)$ 
by an angle $\theta_o$ about the $y$-axis followed 
by rotation through an angle $\phi_o$ about the $z$-axis. 
%smooth and $C^1$ on the 2-sphere. 
%It also produces the best correspondence with the 
%quantum expectation values under dynamical evolution. 
The variance $\sigma^2$ 
%and the length $|{\bf J}|$ 
is a free parameter of the distribution. Although 
$\sigma^2$ cannot be chosen so that all low order moments 
are satisfied, the choice  
$ \sigma^{-2} =   2 |{\bf J}| $, 
where $ |{\bf J}|^2  =  j(j+1) $, produces a reasonable 
compromise, as discussed in \cite{EB00a}.

\section{Dynamical Behaviour of Probability Distributions}

%These classical effects can not be examined directly at the level of 
%the Liouville density since this 4-dimensional phase space 
%can not be conveniently 
%visualised and in quantum mechanics 
%We first consider the signatures of chaos in quatum states and their 
%classical counter-parts by directly examining the time-dependence of the 
%discrete marginal classical distributions (\ref{eqn:pclz}) 
%and (\ref{eqn:pcjz}) and the corresponding quantum 
%ones (\ref{eqn:plz}) and (\ref{eqn:pjz}). 

In the case of a mixing classical system, initial densities 
with non-zero measure are expected to spread in an increasingly 
uniform manner throughout the accessible phase space. 
The term {\em uniform} is meant to apply specifcally 
in a coarse-grained sense. For some simple maps, such as the baker's map, 
it is possible to show that this rate of relaxation to the equilibrium 
configuration occurs exponentially with time \cite{Dorfman}. 

The spin map we consider (\ref{eqn:map}) is not {\em mixing} on the 
accessible classical manifold ${\mathcal P}$, but has {\em mixed} dynamics:  
depending on the system parameters, 
the surface ${\mathcal P}$ can generally  
be decomposed into 
regions of regular dynamics and a connected region of chaotic dynamics 
\cite{EB00a}.  In parameter regimes 
that are predominantly chaotic, we expect 
behaviour on ${\mathcal P}$ that approximates that of a mixing system. 
In particular, initially localised Liouville densities 
should relax close towards the 
microcanonical measure at an exponential rate, on average. 
%In parameter regimes of mixed dynamics, we expect a
%classical density initially localised in the chaotic region 
%to spread uniformly only throughout that region. 
In this section we demonstrate that these signatures of chaos 
are exhibited also by the quantum dynamics. Most striking is the 
degree of similarity between the quantum and classical 
behaviours even in regimes with classically 
mixed dynamics.  

We are interested in the behaviour of quantum probability distributions 
that are associated with measurements of 
classical dynamical variables. The quantum probability distribution 
associated with the classical observable $L_z$ is given by, 
\begin{equation}
\label{eqn:plz}
P_{L_z}(m_l) =  
\langle \psi(n) | \; R_{l,m_l} \; | \psi(n) \rangle 
= {\mathrm Tr} \left[ \; | l,m_l \rangle \langle l,m_l 
| \rho^{(l)}(n)  \; \right], 
\end{equation}
where,
\begin{equation}
R_{l,m_l} =  1_s \otimes | l, m_l \rangle \langle l, m_l | 
\end{equation}
is a projection operator onto the eigenstates of $L_z$, and 
\begin{equation}
\label{eqn:redl}
\rho^{(l)}(n)= {\mathrm Tr}^{(s)} \left[ \; | \psi(n) \rangle 
\langle \psi(n) | \; \right],  
\end{equation}
is the reduced state operator for 
the spin ${\bf L}$ at time $n$ and ${\mathrm Tr}^{(s)}$ denotes 
a trace over the factor space ${\mathcal H}_s$. We have written 
out the explicit expression (\ref{eqn:plz}) to emphasize 
that the probability of obtaining each $m_l$ value is associated with 
a projector onto a subspace of the {\em factor} space ${\mathcal H}_l$. 
%corresponding to the spin ${\bf S}$. 

For reasons related to this fact (which we will make clear in later sections), 
we are also interested in examining the 
probability distributions associated with components of 
the {\it total} angular momentum ${\bf J} = {\bf S} + {\bf L}$.  
%The domain of this operator is the full Hilbert space and 
%is thus described by a state which remains pure throughout 
%the time-evolution. 
%Since the full Hilbert space ${\mathcal H}$ is obtained from a direct sum 
%of irreducible representations for $j$ values varying from $l-s$ to $l+s$, 
The probability of obtaining a given $m_j$ value upon measurement of 
$J_z$ is given by, 
\begin{equation}
\label{eqn:pjz}
P_{J_z}(m_j) = \sum_{m_s} \! 
| \langle \psi(n) |  s, l,m_s,m_j - m_s \rangle |^2,   
\end{equation}
where $|  s, l,m_s,m_j - m_s \rangle$ is an element of the orthonormal basis 
(\ref{eqn:basis}).
The probability $P_{J_z}(m_j)$ 
%of obtaining each $m_j$ value 
is associated with a projector onto a   
subspace of the {\em full} Hilbert space ${\mathcal H}$. 
The dimension of each subspace  
is given by the number of pairs $(m_s,m_l)$ that yield 
a given value of $m_j = m_s + m_l$.

The classical probability distributions associated with dynamical variables 
are obtained by partial integration over the accessible phase space. 
In the case of $L_z$, the continuous marginal distribution is given by, 
\begin{equation}
\label{eqn:pcclz}
	P(L_z) = \int \! \int \! \int dS_z d\phi_s d\phi_l 
\; \rho_c(S_z,\phi_s,L_z,\phi_l), \nonumber
\end{equation}
where for notational convenience 
we have suppressed reference to the time-dependence. 
The marginal probability distribution for the total spin component $J_z$ 
is obtained by integration subject to the constraint $S_z + L_z = J_z$,   
\begin{equation}
\label{eqn:pccjz}
	P(J_z) = \int \! \int \! \int \! \int dS_z d\phi_s d L_z d\phi_l 
\; \rho_c(S_z,\phi_s,L_z,\phi_l) \; \delta(S_z + L_z - J_z). \nonumber
\end{equation}
%In order to compare these classical marginal probability 
%distributions directly with the 
%corresponding discrete quantum distributions, given by (\ref{eqn:plz}) 
%and (\ref{eqn:pjz}), 
%we consider discretized versions of (\ref{eqn:pcclz}) and (\ref{eqn:pccjz})   
%The classical coarse-graining that we adopt is designed to 
%mimick the intrinsic discretization of the quantum 
%probability distributions. 
These classical distributions are continuous, though their quantum 
counter-parts are intrinsically discrete. 
To construct a meaningful quantum-classical comparison 
it is useful to discretize the classical distrbutions 
by integrating the continuous probabilities 
over intervals of width $\hbar =1$ centered on 
the quantum eigenvalues. 
In the case of the component $L_z$, 
the quantum probability $P_{L_z}(m_l)$ is then associated with 
%we compare the quantum probabilities $P_z(m_l)= \langle \psi(n) | 
%m_l \rangle \langle m_l | \otimes 1_s  \psi(n)\rangle = 
%{\mathrm Tr} (|m_l \rangle \langle m_l | \rho_l(n))$ 
the classical probability of finding $L_z$ in the 
interval $[m_l-1/2,m_l+1/2]$. This is given by  
%. Therefore we construct 
%a discrete classical probability distribution 
\begin{equation}
\label{eqn:pclz}
%\begin{eqnarray}
	P_{L_z}^c(m_l)  = \int_{m_l - 1/2}^{m_l + 1/2} P(L_z). 
%\nonumber 
%\end{eqnarray}
\end{equation}
%with $P(L_z)$ given by (\ref{eqn:pclz}). 
% which is determined from the continuous marginal distribution 
Similarly, in the case of $J_z$, we compare each quantum $P_{J_z}(m_j)$ with 
the {\em discrete classical} probability, 
\begin{equation}
\label{eqn:pcjz}
P_{J_z}^c(m_j) = \int_{m_j - 1/2}^{m_j + 1/2} P(J_z).
\end{equation}
%with $P(J_z)$ given by (\ref{eqn:pcjz}). 

In the following discussion of the numerical results we will 
emphasize that, for chaotic states, 
the steady-state shape of the quantum and classical 
distributions should be compared 
with the corresponding set derived from the microcanonical state. 
Our model  
is non-autonomous, but the spin magnitudes are conserved. The 
appropriate {\em classical} microcanonical measure is  
a constant on the accessible manifold ${\mathcal P} = S^2 \times S^2$.  
This follows 
from the usual equilibrium hypothesis that all accessible microstates 
are equiprobable, where equiprobability is defined 
with respect to the invariant measure (\ref{eqn:measure}). 
This microcanonical density projected onto the $L_z$-axis 
produces the discrete, flat distribution, 
\begin{equation}
\label{eqn:pmclz}
P_{L_z}^{mc}(m_l)=(2l+1)^{-1}. 
\end{equation}
However, projected along $J_z$, 
the microcanonical distribution is not flat, but has a tent-shape, 
\begin{eqnarray}
\label{eqn:pmcjz}
P_{J_z}^{mc}(m_j) & =  &  \frac{l+s+1-|m_j|}{(2s+1)(2l+1)} 
\;\;\; {\mathrm for} \;\;\;  |m_j| \geq l-s  \nonumber \\ 
& = &  \frac{1}{2l+1}  \;\;\; {\mathrm for} \;\;\; |m_j| \leq l-s. 
%&  &  \frac{l+s+1-m}{(2s+1)(2l+1) } \;\;\; 
\end{eqnarray}
%This functional form is determined by  
%assuming that each of the discrete $m_s$ and $m_l$ intervals are 
%equally likely to be occupied by the subsystem spins,  
%and then counting the number of ways in which 
%these subsystem values combine to produce a given $m_j = m_s + m_l$. 
In quantum mechanics, the equiprobability hypothesis 
implies that the appropriate microcanonical state 
is an equal-weight mixture. This microcanonical state, sometimes 
called a random state, is  
proportional to the identity in the full Hilbert space 
${\mathcal H} = {\mathcal H}_s \otimes {\mathcal H}_l$. 
It produces the same projected 
microcanonical distributions, 
{\it i.e}.\ (\ref{eqn:pmclz}) for $L_z$ 
and (\ref{eqn:pmcjz}) for $J_z$, as the classical microcanonical state. 
%$|s,l,j,m_j\rangle$ contributing to (\ref{eqn:plz}) and (\ref{eqn:pjz}) 
%are equally occupied. and then summing over 
%each of the $j$ representations which produce a given $m_j$ value.  

\subsection{Mixed Regime Chaos}

We consider first a classical parameter regime ($\gamma = 1.215$, 
$r=1.1$, and $a=5$) for which the kinematically 
accessible phase space ${\mathcal P}$ is highly mixed. 
%a dynamical regime for which the kinematic surface 
%${\mathcal P}$ is almost evenly mixed. 
%For classical parameters $\gamma = 1.215$, 
%$r=1.1$, and $a=5$, the anti-parallel fixed points $(\pm S_z,\mp L_z)$ 
%are unstable, but the parallel fixed points $(\pm S_z, \pm L_z)$  
%are stable (and thus surrounded by regular zones) \cite{EB00a}. 
%We sampled ${\mathcal P}$ using $3 \times 10^4$ initial conditions 
%and found that the chaotic zone appears to be connected 
The chaotic region appears to be connected (all chaotic 
initial conditions have the same 
largest Lyapunov exponent $\lambda_L = 0.04$) and covers 
about half of the kinematic surface. A projection of only the 
chaotic initial conditions onto the plane spanned by $S_z$ and $L_z$ 
reveals large regular islands surrounding the stable parallel fixed points 
$(\pm S_z, \pm L_z)$, with chaotic regions spreading out from 
the unstable anti-parallel fixed points $(\pm S_z,\mp L_z)$.  
%with the  chaotic initial conditions clustered near the anti-parallel 
%fixed points. 
A similar projection of the regular initial conditions 
shows points not only clustered about the parallel fixed points but 
also spread along the axis $\tilde{S}_z = \tilde{L}_z$. 
%This implies that initial conditions corresponding to 
%parallel spins along any direction produce predominantly regular 
%dynamics. 

\begin{figure}
%\begin{center}
\hspace{0.75in} 
\input{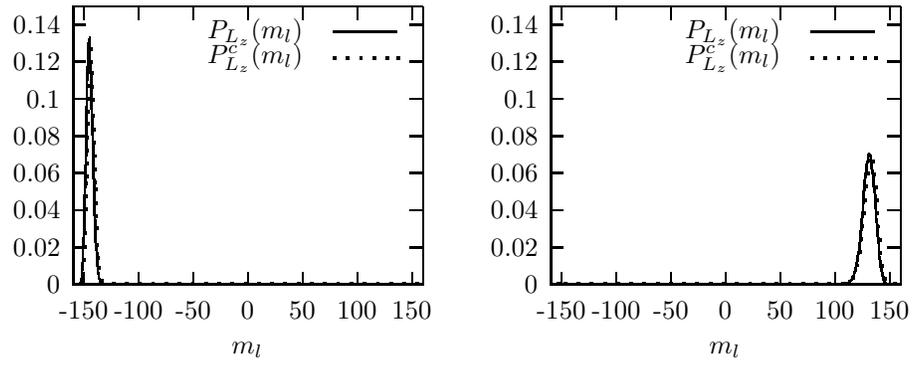}
\vspace{0.2in}
\caption{
Quantum and classical probability distributions for $L_z$ with $l=154$  
in chaotic zone of mixed regime ($\gamma=1.215,r=1.1,a=5$). 
The dots are visible because they are shifted to 
the right by half of their width. 
The figure on the left is the initial state ( $n=0$) 
and that on the right is at time-step $n=6$.
%and initial condition $\vec{\theta}(0) = (20^o,40^o,160^o,130^o)$. 
%showing that the initial distributions remain well localised and very 
%well-matched.
}
\label{g1.215.plz.140.00.06}
%\end{center}
\end{figure}

\begin{figure}
%\begin{center}
\hspace{0.75in} 
\input{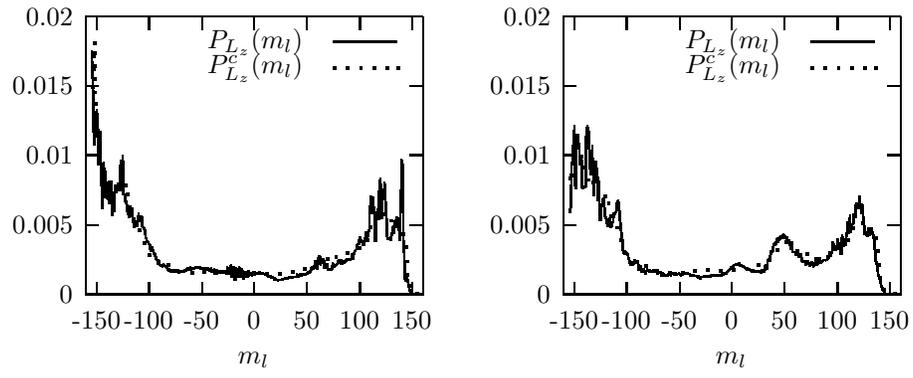}
\vspace{0.2in}
\caption{Same as Fig.\ \ref{g1.215.plz.140.00.06} but for time-steps 
$n=99$ on the left and $n=100$ on the right.  Both quantum and 
classical distributions have reached the system dimension 
and are relaxing towards equilibrium.}
\label{g1.215.plz.140.100}
%\end{center}
\end{figure}

\begin{figure}
%\begin{center}
\hspace{0.75in} 
\input{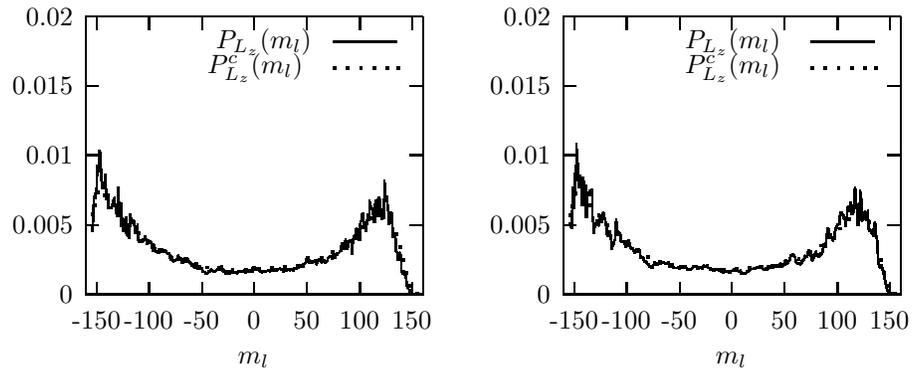}
\vspace{0.2in}
\caption{Same as Fig.\ \ref{g1.215.plz.140.00.06} but for $n=199$ on the 
left and $n=200$ on the right.
The quantum distribution is fluctuating about a classical steady-state.  
}
\label{g1.215.plz.140.200}
%\end{center}
\end{figure}

We now consider the time-evolution of quantum and classical states 
concentrated in the chaotic zone near one of the unstable anti-parallel 
fixed points,  with initial centroids directed along 
$\theta(0)=(20^o,40^o,160^o,130^o)$. 
The quantum dynamics are calculated using quantum numbers $s=140$ 
and $l=154$.  
As shown in Fig. \ref{g1.215.plz.140.00.06}, at early times 
both the quantum distribution $P_{L_z}(m_l)$ (solid line) 
and the corresponding classical distribution $P_{L_z}^c(m_l)$ (dots)
remain well-localised. Their initial differences are not 
distinguishable on the scale of the figures. 
(The dots are shifted to the right by half of their width.) 
By time-step $n=20$ both quantum and classical 
distributions have broadened to the system dimension and 
begin to exhibit noticeable differences. 
As shown in Fig.\ \ref{g1.215.plz.140.100}, 
around $n=100$ the distributions have begun to settle close 
to an equilibrium shape. 
%steps $n=99$ and $n=100$. 
In Fig.\ \ref{g1.215.plz.140.200} the successive 
time steps $n=199$ and $n=200$ show that,  although both the 
quantum and classical distributions have relaxed very close 
to the same equilibrium distribution, the quantum distribution 
exhibits rapidly oscillating fluctuations about the classical 
steady-state. 
%Consequently these fluctuations would be very difficult to observe 
% coarse-grained. 

\begin{figure}
%\begin{center}
\hspace{0.75in} 
\input{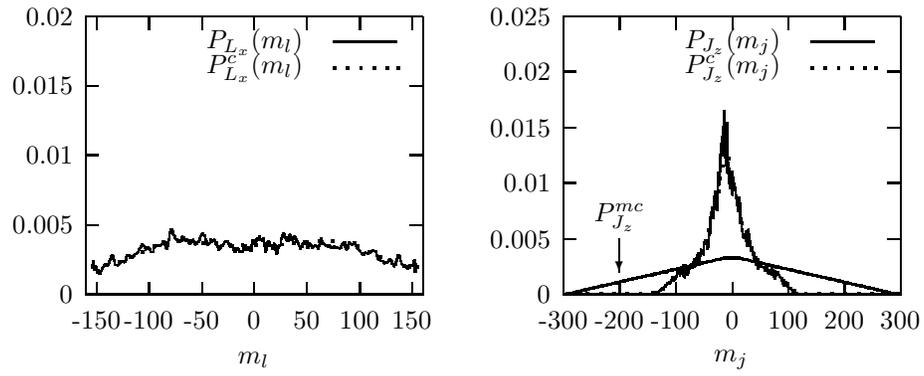}
\vspace{0.2in}
\caption{
Same as in the previous figure, but for 
$P_{L_x}(m_l)$ on the left and 
$P_{J_z}(m_j)$ on the right, at time-step $n=200$.  
Both $P_{J_z}(m_j)$ and $P_{J_z}^c(m_j)$ 
are localised relative to the 
projected microcanonical distribution $P_{J_z}^{mc}(m_j)$.   
}
\label{g1.215.plx.pjz.140}
%\end{center}
\end{figure}

Both the quantum and classical 
equilibrium distributions (projected along $L_z$) 
show significant deviation from the 
microcanonical distribution  (\ref{eqn:pmclz}).
This is also true of the distribution projected 
along $L_x$, which has a different non-uniform 
equilibrium distribution than 
that observed when projecting onto $L_z$ (see the 
left box of Fig.\ \ref{g1.215.plx.pjz.140}). 
Uniform marginal distributions would be expected 
if the classical mapping was mixing, in which case arbitrary 
initial densities (with non-vanishing measure) would relax to  
the microcanonical distribution. Since the accessible 
kinematic surface has large KAM surfaces in this parameter 
regime, the coarse-grained {\em classical} equilibrium 
distributions are not expected to be flat.  
An unexpected feature of the results is the observation 
that the shape of the equilibrium 
{\em quantum} distributions so accurately reflects the 
details of the KAM structure  in the classical phase space. 
This feature is most striking in the case of the distributions 
projected along $J_z$ (see the right box 
in Fig.\ \ref{g1.215.plx.pjz.140}).
%The unique shape 
%of the $P_{L_x}(m_l)$ steady-state distribution  
%is also closely matched by the quantum distribution along that axis 
%(modulo the irreducible quantum fluctuations).
%In the case of the total spin ${\bf J} = {\bf S} + {\bf L}$, 
%shown on the right in Fig.\ \ref{g1.215.plx.pjz.140}, the 
The steady-state quantum and classical 
probability distributions $P_{J_z}(m_j)$ and $P_{J_z}^c(m_j)$ 
are both sharply peaked about $m_j = 0$. 
This equilibrium shape is much more sharply peaked 
than the tent-shape of the 
projected {\em microcanonical} distribution, 
$P_{J_z}^{mc}(m_j)$, given by (\ref{eqn:pmcjz}) and 
also plotted in the right box in Fig.\ \ref{g1.215.plx.pjz.140}. 
The important point is that the 
additional localization of the {\em quantum} distribution 
can be understood from a standard 
fixed-point analysis of the {\em classical} map \cite{EB00a}:    
the presence of KAM surfaces arising due to the stability 
of the parallel fixed points 
prevents the chaotic classical spins 
from aligning in parallel along the $z$-axis. 
%The presence of these KAM surfaces can be predicted 
%from a fixed-point of the classical map (\ref{EB00a}). 
%Though we have 
%confirmed analytically the existence of such KAM surfaces surrounding
%only the north and south poles from our analysis of the trivial fixed 
%points \cite{EB00a}, the numerical analysis, mentioned above, 
%confirms that initial conditions 
%with $\tilde{S}_z \simeq \tilde{L}_z$ 
%are found to be predominantly regular. 
%This Once again we emphasize that 
Most remarkably, we find that 
the steady-state quantum distributions accurately reproduce this  
parameter-dependent structure of the mixed classical phase space 
even for much smaller quantum numbers. We examine 
how the accuracy of this correspondence 
scales with the quantum numbers in section \ref{sect:scaling}.

\subsection{Regime of Global Chaos}

If we hold $a=5$ and $r=1.1$ fixed and increase the coupling strength 
to the value $\gamma=2.835$,  
then all four of the fixed points mentioned above 
become unstable \cite{EB00a}. 
%By directly sampling the accessible kinematic surface 
%${\mathcal P}$ with $3 \times 10^4$ 
%initial conditions we find that 
Under these conditions less than $0.1\%$ of the surface ${\mathcal P}$
is covered with regular islands; the remainder of the surface 
produces a connected chaotic zone with 
largest Lyapunov exponent $ \lambda_L = 0.45 $.  We will sometimes 
refer to this parameter regime as one of {\em global chaos} since the 
kinematically accessible phase space is predominantly chaotic.  

\begin{figure}
\hspace{0.75in} 
%\begin{center}
\input{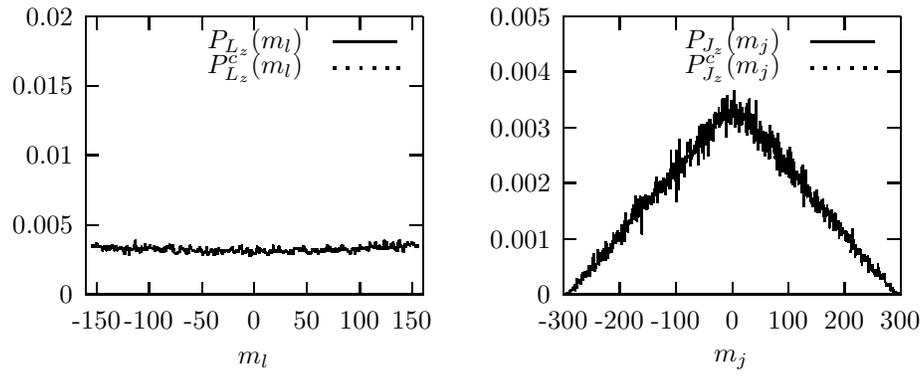}
\vspace{0.2in}
\caption{
The equilibrium shapes of $P_{L_z}(m_l)$ 
and $P_{J_z}(m_j)$ at time-step $n=50$ with $l=154$ 
for a state launched in the global chaos regime 
($\gamma=2.835,r=1.1,a=5$). 
The quantum distributions exhibit small rapidly 
oscillating fluctuations about the 
projected microcanonical distributions. 
The classical distributions
% $P_z^{mc}(m_j)$. 
% $P_z^c(m_j)$, 
are not visible since the points 
lie within the 
fluctuating quantum data.  
}
\label{g2.835.plz.pjz.140.50}
%\end{center}
\end{figure}

The dynamics of the classical and quantum distributions are much simpler 
in this regime. We find that initially localised 
distributions, launched from 
arbitrary initial conditions, relax to the microcanonical 
distribution on a very short time-scale. To demonstrate this, 
we consider the dynamics of 
an initial quantum state with $s=140$ and $l=154$, and a 
corresponding classical 
density, launched from $\theta(0) = (20^o, 40^o, 160^o, 130^o)$. 
Though the initial distributions are the same as in the mixed regime,  
by time-step $n \simeq 6$ 
the quantum and classical distributions 
have already spread to the system dimension and begin 
to exhibit noticeable 
differences. By time-step $n \simeq 12$
both distributions have relaxed 
very close to the microcanonical distributions.  
We plot the equilibrium quantum and classical 
projected distributions $P_{L_z}(m_j)$ and $P_{J_z}(m_j)$ 
in Fig.\ \ref{g2.835.plz.pjz.140.50} for time-step $n=50$.
The projected {\em classical} distributions are early indistinguishable 
from the microcanonical forms, $P_{L_z}^{mc}(m_j)$ and $P_{L_z}^{mc}(m_j)$,  
and  the {\em quantum}
distributions again exhibit small fluctuations 
about the classical distributions. 
We have found that these equilibrium 
quantum-classical differences asymptote to  
a non-vanishing minimum when the measure of KAM surfaces becomes negligible. 
These minimum quantum fluctuations 
reflect characteristic deviations from the microcanonical state that arise 
because the equilibrium quantum state is pure, 
whereas the microcanonical state corresponds to a random mixture. 

%classical equilibrium. 
%The quantum distribution 
%has relaxed to the tent-shaped microcanonical 
%distribution (\ref{eqn:mcjz}), as shown in 
%Fig.\ \ref{g2.835.pjz.peq.294.n50}, corresponding to $n=50$. 

%\input{HDeq.tex} 
\section{Rates of Relaxation to Equilibrium}

In order to characterize the time-scale of relaxation 
to equilibrium it is convenient to study the time-dependence 
of a scalar 
measure that is sensitive to deviations from the equilibrium 
state.  A conventional indicator of this rate of 
approach to equilibrium is the coarse-grained entropy,  
\begin{equation}
\label{eqn:hb}
	H = - \sum_i P_i \ln P_i. 
\end{equation} 
%considered by Boltzmann in his well-known H-theorem \cite{Dorfman}.  
%That theorem was developed to justify the postulates of equilibrium 
%statistical mechanics. 
Here the $\{P_i\}$ stand for the quantum probabilities associated 
with projectors onto some basis of microstates ({\it e.g}.\ the projected 
distributions discussed in the previous section). 
The sum (\ref{eqn:hb}) is a standard measure of the 
information contained in a probability distribution 
and is sometimes called the Shannon entropy.  

The Shannon entropy has a number of useful properties. 
First, unlike the von Neumann entropy ${\mathrm Tr} [ \rho \ln \rho]$, 
the Shannon entropy is basis-dependent. It reduces to the von Neumann 
entropy if the `chosen' basis diagonalizes the state operator. 
However, this basis, or, more precisely, the set of projectors onto the 
(time-dependent) spectral decomposition of the state operator, 
does not necessarily correspond to a 
set of classically meaningful observables.   
Our main interest is to examine correspondence 
at the level of classical dynamical variables, so we consider 
probabilty distributions associated with projectors onto the 
eigenstates of classically well-defined operators.  
The classical counterparts to these probability distributions 
are associated  with some fixed partioning of the phase space  
into cells of width $\hbar$ along the axes of the associated dynamical 
variable. 
%which presumes 
%some partioning of cells in phase space that is fixed externally 
%({\it e.g}.\ the lab frame). These cells are chosen to correspond 
%to the set of projectors that diagonalize 

Second, whereas the von Neuman entropy 
of the total system is constant in time 
(${\mathrm Tr} [ \rho \ln \rho]=0$ since 
$\rho = |\psi \rangle \langle \psi |$), 
the basis-dependent Shannon entropy 
may have time-dependence even if the quantum state is pure. 
Thus (\ref{eqn:hb})  may be applied to examine the rate of relaxation 
of either {\em pure} or {\em mixed} quantum states. 
It is in this sense that we use the term {\em relaxation},  
although the time-evolution is unitary in the quantum model 
(and volume-preserving in the classical model). 

Given some fixed partioning of the phase space, if a 
classical state remains evenly spread through the phase space cells 
it occupies, and  
spreads through the phase-space exponentially with time, then 
an entropy like (\ref{eqn:hb}) 
should grow linearly with time. 
%In a simple approximation we may take \cite{Peres96}. 
In this section we show that this argument 
holds approximately also for quantum states  
launched from a classically chaotic region of phase space. 
The actual rate of relaxation of the quantum states  
is accurately predicted by the classical entropy even for small 
quantum numbers. 
%Consequently the Shannon 
%entropy can serve as a convenient indicator of quantum chaos even if 
%the quantum state is pure. 

%For the Baker's map, an idealised hyperbolic system, 
%it can be shown that there exists a coarse-graining 
%for which the Shannon entropy 
%grows linearly in time, on average, at rate given approximately 
%by the sum of the positive Lyapunov exponents. 
%Such a rule can be expected to hold, approximately, for 
%more complex classical systems with a mixed phase space. 

\begin{figure}
\begin{center}
%\hspace{0.75in} 
\input{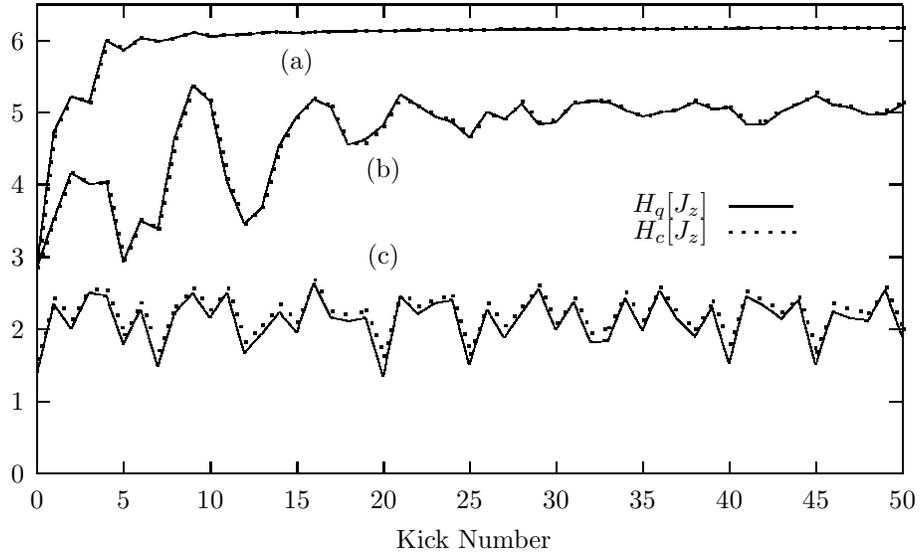}
\vspace{0.2in}
\caption{
Comparison of the quantum and classical entropies 
$H[J_z] = - \sum_{m_j} P_{J_z}(m_j) \log  P_{J_z}(m_j)$ 
for $s=140$ and $l=154$ in 
(a) regime of global chaos ($\gamma=2.835$); (b) chaotic zone of 
the mixed regime ($\gamma=1.215$); 
(c) regular zone of the mixed regime ($\gamma=1.215$).
}
\label{HJz.294}
\end{center}
\end{figure}

We demonstrate this behaviour by first considering the 
quantum entropy $H_q[J_z]$ of 
the probabilities associated with the eigenvalues $m_j$ 
of $J_z$, {\it i.e}.\ the probabilities defined in (\ref{eqn:pjz}).
% and (\ref{}), respectively. 
The corresponding classical entropy, $H_c[J_z]$, is calculated using 
the discrete classical probabilities (\ref{eqn:pcjz}).  
%that are associated with a set of intervals of width 
%$\hbar=1$ along the $J_z$ axis.
In Fig.\ \ref{HJz.294} we compare the 
time-development of the quantum and classical entropies 
% for the probability distribution 
%of the total system spin $P_z(m_j)$
using quantum numbers $s=140$ and $l=154$. 
For these quantum numbers,  
%and using (\ref{eqn:pmcjz}), 
%$H[J_z]$ has 
the microcanonical ({\it i.e}.\ maximum) value of 
the entropy is $H^{mc}[J_z]=6.2$. 
In case (c), corresponding to 
a regular zone of the mixed regime 
($\theta(0) =(5^o,5^o,5^o,5^o)$,$\gamma=1.215$), we actually see 
the greatest amount of difference between the quantum 
($H_q[J_z]$) and classical ($H_c[J_z]$) entropies. 
$H_q$ exhibits a quasi-periodic 
oscillation about its initial value whereas for $H_c$ these oscillations 
eventually dampen. 
For smaller quantum numbers, and thus broader initial states,  
$H_c$ dampens much more rapidly although $H_q$ continues to exhibit 
a pronounced quasi-periodic behaviour.  
% before relaxing to a constant well above the 
%microcanonical minimum. 
In case (b), 
with initial centroid $\theta(0)= (20^o,40^o,160^o,130^o)$ 
set in a chaotic region of the equally mixed regime, 
both $H_q$ and $H_c$ oscillate about an initially increasing average
before relaxing towards a constant value that lies well below 
the microcanonical maximum $H^{mc}[J_z] = 6.2$. 
This saturation away from the maximum is expected  
in the classical model since a large fraction of the kinematic surface 
is covered with regular islands and remains inaccessible.  
In case (a), corresponding  
to the regime of global chaos ($\gamma =2.835$) 
and with the same initial state as (b), 
%$\theta(0)= (20^o,40^o,160^o,130^o)$, 
the entropies are nearly identical. Both grow much more quickly 
than in the mixed regime case, roughly linearly, 
until saturating very near the maximum value. 

\begin{figure}
\begin{center}
%\hspace{0.75in} 
\input{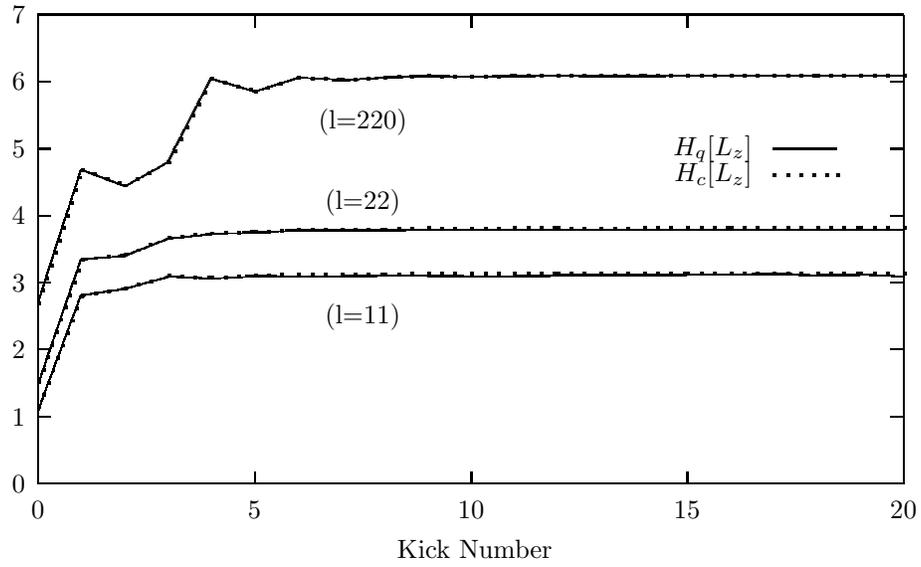}
\vspace{0.2in}
\caption{
Comparison of the quantum and classical subsystem entropies 
$H[L_z] = - \sum_{m_j} P_{L_z}(m_l) \ln  P_{L_z}(m_l)$ 
for increasing system sizes in the global chaos regime of 
Fig.\ \ref{HJz.294}.
}
\label{Hent.Lz.11.22.220}
\end{center}
\end{figure}

The quantum entropy is very well approximated by its  
classical counterpart also for smaller quantum numbers. 
In Fig.\ \ref{Hent.Lz.11.22.220} we display the growth rates 
of the quantum and classical entropies of the probabilty 
distributions associated with the observable $L_z$ for 
three sizes of quantum system ($l=11$, $l=22$, $l=220$) using 
the same parameters and initial condition as for 
data-set (a) in Fig.\ \ref{HJz.294}. 
In each case the quantum entropy is essentially identical 
to the corresponding classical entropy. The initial rate 
of growth is similar in each case, roughly linear, 
and of order the Lyapunov exponent, $\lambda_L =0.45$.  

These results extend previous work demonstrating the 
that the widths of quantum states grow exponentially with 
time, on average, until saturation at the system dimension 
\cite{Fox94b,Ball98,EB00a}. Modulo the small quantum 
fluctuations, for both quantum and classical 
models we find that the subsequent relaxation to an 
equilibrium  configuration occurs on the time-scale,  
\begin{equation}
t_{rel} \sim t_{sat} + {\mathcal O}(\lambda_L^{-1}),  
\end{equation}
where $t_{sat} \simeq \lambda_{w}^{-1} \ln l$  estimates 
the time it takes the initial coherent state 
to reach the system dimension.  
The exponent $\lambda_{w}$ is the exponent governing  
the growth rate of the state width \cite{EB00a}.   
The last term ${\mathcal O}(\lambda_L^{-1})$ approximates  
the additional time-required for the state 
to become more or less uniformly spread 
over the accessible phase space. 
%the exponential spreading of 
%the initially localised distributions.  
%and is of order the largest Lyapunov exponent $\lambda_L$.   
%This exponential spreading 
%saturates when the distributions reaches the system dimension, on 
%the time-scale $t_{sat} \simeq \lambda_{w}^{-1} \log l$. 
In predominantly chaotic regimes we have found that  
$\lambda_w \simeq \lambda_L$, 
though in mixed regimes $\lambda_w$ is generally a few times 
larger than the largest Lyapunov exponent. 
%Its magnitude  can be estimated more precisely 
%from the actual growth rate of the classical variance.  
%The explicit dependence on the quantum number $l$ arises from taking 
%the ratio of the standard deviation of the 
%initial distribution to the accessible system dimension. 
%Consequently this log estimate applies only when the quantum states are 
%maximally localised for a given set of quantum numbers. 
%This occurs sooner 
%for projected distributions than for full 
%distributions defined on the accessible manifold \cite{Dorfman}.

%\input{Dmc.tex} 
\section{Time-dependence of Quantum-Classical Differences} 

Before examing the scaling of quantum-classical differences 
with increasing quantum numbers, it is useful to determine 
first their time-domain characteristics 
under the different types of classical behviour. 
Previous work has shown that quantum-classical differences 
for low-order moments, though initially 
small, grow exponentially with time when the classical motion is chaotic  
\cite{Ball98,Ball00} until the states approach the system size 
\cite{EB00a}. On this saturation 
time-scale those quantum-classical differences reach   
their maximum magnitude, 
but surprisingly this maximum was small, ${\cal O}(\hbar)$. 
More specifically, it did not scale with the quantum numbers. 
Of course, two distributions can be altogether different even when   
the differences between their means and variances are quite small, 
and therefore it is useful to examine the differences between 
the quantum and classical states in a more sensitive way. 
%In particular, information about differences in the fine structure 
%may be completely lost  

In this section we examine the time-dependence of 
%a much more sensitive measure of 
%quantum-classical differences by 
bin-wise deviations 
between the quantum and classical probability distributions.  
For the observable $L_z$ 
this indicator takes the form,  
%using  the difference measure: 
%eviations be a bin-wise 
%comparison of the probability  distributions: 
\begin{equation}
\label{eqn:siglz}
\sigma[L_z] = \sqrt{ {1 \over (2l+1)} \sum_{m_l = -l}^l  
[ P_{L_z}(m_l) - P_{L_z}^c(m_l) ]^2 }.
\end{equation}
This standard deviation estimates typical 
quantum-classical differences on the scale of $\hbar$ along 
the $L_z$-axis. Each interval is centered on a quantum eigenvalue.
The $ P_{L_z}^c(m_l)$ correspond to 
a measurement, or coarse-graining, of the classical density 
on an extremely fine-scale. 
% of a classical dynamical variable. 
%Its magnitude 
%should be compared with a typical value of either distribution, 
%$P(m_l) \sim (2l+1)^{-1} \simeq  3 \times 10^{-3}$ for $l=154$. 

\begin{figure}
\begin{center}
\input{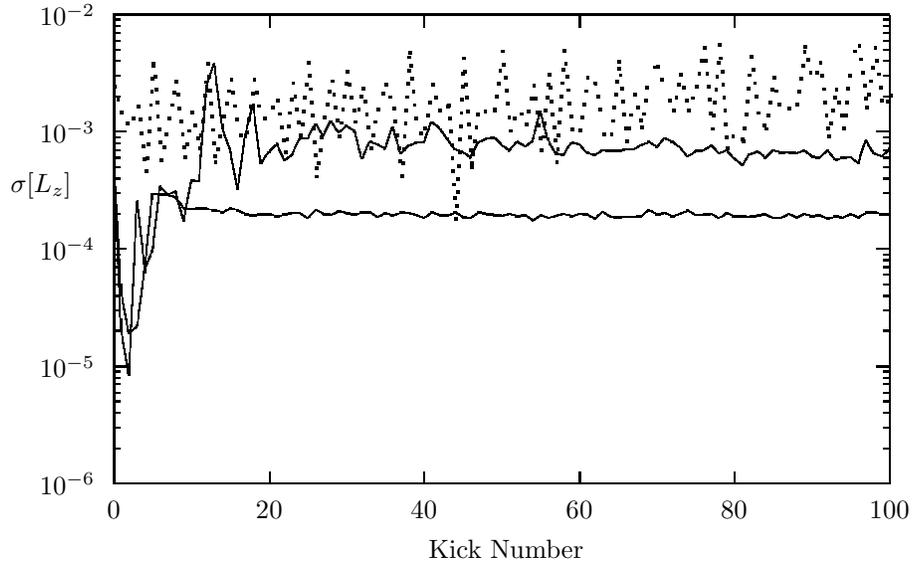}
\vspace{0.2in}
\caption{
Time dependence of the standard deviation $\sigma[L_z]$ of 
quantum-classical differences (\ref{eqn:siglz})
for states launched from a regular zone (dotted line) 
of the mixed regime ($\gamma=1.215$), from a chaotic zone of the same 
mixed regime (middle solid line), 
and from the regime of global chaos (lower solid line,$\gamma=2.835$).  
The initial discrepancy 
is relatively large, but quickly decreases, and then increases until 
reaching an asymptotic equilibrium value. This occurs more slowly for 
the mixed regime case, for which the asymptotic value is also larger. 
In all cases $s=140$ and $l=154$. 
}
\label{D2.Lz.154}
\end{center}
\end{figure}

In Fig.\ \ref{D2.Lz.154} we examine the time dependence of $\sigma[L_z]$ for 
the same three classical sets of parameters and initial conditions 
displayed in Fig.\ \ref{HJz.294}. 
The initial value of $\sigma[L_z]$ is generally 
not zero since it is not possible to match all the marginal distributions 
exactly in the case of the SU(2) coherent states \cite{EB00a}. 
The actual magnitude of the initial discrepancy depends 
on the angle between the axis of measurement, {\it e.g}.\ $L_z$, and 
the direction of polarization of the initial state.  
%As was evident from the earlier direct comparisons 
%of the probability distributions, this is simply a reflection of 
%the fact that for a carefully constructed classical density 
%the distribution differences are generally small for both 
%regular and chaotic states over long times. 
For both chaotic states the differences initially decrease from their 
angle-dependent value  
and then increase until saturation at a steady-state value. 
This steady-state value is  reached much later 
in the mixed regime (upper solid line), than in 
the global chaos regime (lower solid line).  
It occurs on the time-scale, $t_{rel}$, on which    
the underlying distributions have reached their steady-state 
configurations (modulo the quantum fluctuations). 

%The typical magnitude of the quantum-classical differences in 
%the steady-state regime is most interesting for reasons 
%discussed in the next section. 
%Perhaps the most interesting feature of the correspondence revealed by 
%this measure is 
As shown in Fig.\ \ref{D2.Lz.154}, the quantum-classical differences are 
actually largest for the regular state (dotted line) of the mixed regime 
($\gamma=1.215$) at both early and late times (relative to the 
relaxation time-scale). 
The steady-state magnitude of the differences for the global chaos 
regime ($\gamma=2.835$) 
is significantly smaller than the typical magnitude 
for the mixed regime. 
%As we will see later, this is 
%probably due to the fact that accessible Hilbert space is 
%much smaller. 
However, for larger values of the classical 
perturbation strength $\gamma$,
this average steady-state magnitude does not decrease further   
(with the quantum numbers held fixed) but has reached 
a non-vanishing minimum. 
The magnitude of the minimum steady-state fluctuations,  
$D_{L_z} \simeq 2 \times 10^{-4}$, should be 
compared with a typical magnitude 
of the quantum and classical distributions, 
$ P_{L_z}(m_l) \simeq 3 \times 10^{-3}$.  
In the following section 
we examine how these fluctuations scale with increasing 
quantum numbers.

%[[Since the total state is pure, the average degree of 
%mixing of the subsystem state  
%(and thus the magnitude of the minimum flucutations)  
%may depend on the dimension of the other factor space.
%Then the magnitude of these minimum quantum fluctuations may be 
%determined by the full Hilbert space dimension, rather than 
%merely the dimension of the factor space.  (In our model the dimensions  
%of the factors space are kept at a fixed ratio.)]]

\begin{figure}
\begin{center}
\input{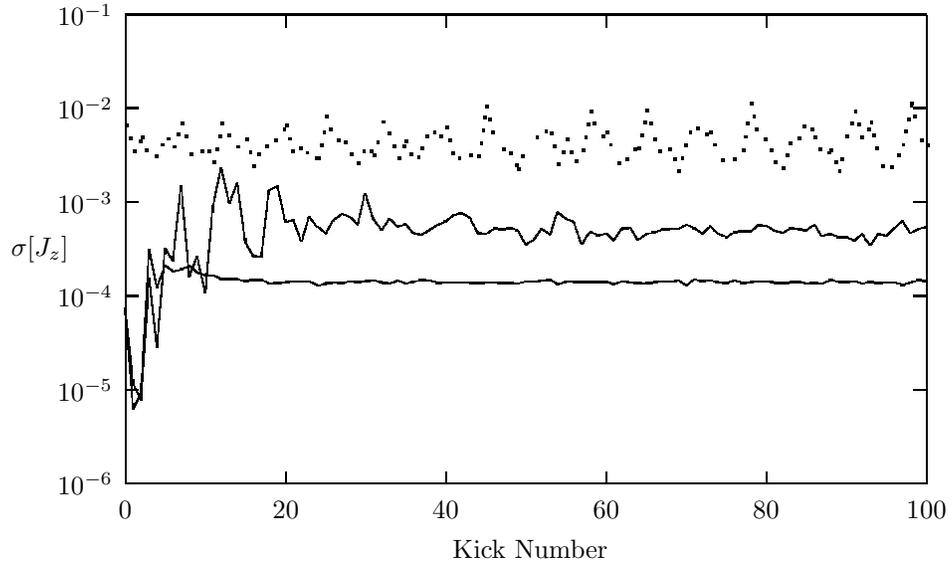}
\vspace{0.2in}
\caption{
Same as Fig.\ \ref{D2.Lz.154} but for the standard deviation of 
quantum-classical differences of the total angular momentum, 
$\sigma[J_z]$, given by (\ref{eqn:sigjz}).
}
\label{D2.Jz.154}
\end{center}
\end{figure}

Above we have considered quantum-classical differences 
for observables (projectors onto subspaces) associated with the  
factor space ${\mathcal H}_l$. 
In this factor space the state is initially pure but 
becomes mixed as a result of dynamical interacions with the 
other subsystem. 
%This effect is analogous to the process of decoherence. 
It is interesting to check if the dynamical 
behaviours of the differences are an artefact of this dynamical mixing.    
Therefore we consider also bin-wise quantum-classical differences 
for an observable ($J_z = S_z + L_z$) that acts non-trivially 
on the full Hilbert space ${\mathcal H} = {\mathcal H}_s 
\otimes {\mathcal H}_l$. The quantum 
state in the full Hilbert space remains pure throughout 
the time-evolution.  
We construct the same standard deviation of the bin-wise 
differences between the quantum and 
classical probability distributions as above, 
\begin{equation}
\label{eqn:sigjz}
\sigma[J_z] = \sqrt{ {1 \over [2(s+l)+1)]} \sum_{m_j}   
[ P_{J_z}(m_j) - P_{J_z}^c(m_j) ]^2 }, 
\end{equation}
where $m_j \in \{l+s, l+s-1, \dots, -(l+s) \}$. 
In Fig.\ \ref{D2.Jz.154} we compare $\sigma[J_z]$ in 
the same three classical regimes examined in Fig.\ \ref{D2.Lz.154}. 
%Differences between quantum and classical observables 
%show similar behaviour although the state is pure and the dynamics 
%are unitary or not. 
Once again the regular state (dotted line) exhibits the 
largest quantum-classical differences, and the differences for 
both chaotic states (middle and lower solid line) 
grow to a steady-state value on the time-scale at which 
the underlying distributions relax to their equilibrium configurations.
As above, the average value of the differences for 
$\gamma=2.835$ (lower solid line) correspond to a 
non-vanishing minimum, that is,  
the average value does not 
noticeably decrease for larger values of $\gamma$. 
The minimum quantum fluctuations are again small 
when compared with the average height of the probability 
distribution, $[2(s+l) +1]^{-1} \simeq 2 \times 10^{-3}$. 
%%%% THIS IS THE STUFF ABOUT ELI'S FLUCTUATIONS: 

For $\gamma \simeq 2.835$, the measure of regular islands 
is already very close to zero and the classical system 
is nearly ergodic on ${\mathcal P}$.  Similarly, the quantum 
state is no longer constrained by any invariant classical 
structures but spreads {\em almost} evenly about the accessible 
Hilbert space. We find that the standard deviations of the 
quantum fluctuations 
that account for the equilibrium quantum-classical differences 
approach a non-vanishing minimum as the classical dynamics 
approach ergodicity on ${\mathcal P}$.
These equilibrium differences can not vanish (for fixed quantum numbers) 
because the total quantum state remains pure under the unitary dynamics,  
whereas the microcanonical equilibrium corresponds to an 
equal-weight mixture.  

%\cite{Page93,Lubkin78}. 
%may correspond to the characteristic 
%fluctuations that arise when the 
%random state , whereas 

\section{Correspondence in the Classical Limit}
\label{sect:scaling}

We now turn to an examination of the classical limit, 
${\mathcal J}/\hbar \rightarrow \infty$, where ${\mathcal J}$ 
is characteristic system action. Since the quantum-classical 
differences grow to their largest values once the states 
have spread to the system size and subsequently fluctuate about this 
magnitude, we will examine the scaling of the differences 
in this late time-domain, that is,  
when the states 
have relaxed close to their equilibrum configurations. 
Moreover, these scaling results will then 
complement previous work 
that has focussed on correspondence at early times \cite{EB00a}, 
in the Ehrenfest regime when the states 
are narrow relative to the system dimensions. 

We wish to determine if the standard deviation of the 
quantum-classical differences (defined in the previous 
section) decreases in magnitude 
with increasing quantum numbers. 
When comparing models with increasing quantum 
numbers, we hold the width of each probability 
bin fixed (at $\hbar=1$). Since the number of bins 
will increase with the quantum number, it follows that  
the height of the probability distribution in a 
given bin will also decrease. Consequently, 
we construct a scale-independent, or relative, measure of 
the bin-wise quantum fluctuations by taking the ratio of the standard 
deviation to the average value of the probability distribution. 
For the observable $L_z$ this takes the form,   
\begin{equation}
\label{eqn:rlz}
R [L_z(n)] =  \frac{D [L_z(n)]}{{\overline P}_{L_z}}  = N_l \; \sigma[L_z(n)]. 
\end{equation}
where the average value $  {\overline P}_{L_z} = 1/(2l+1) = 1/N_l$. 
%(away from the classical distribution)  
%relative to the magnitude of the classical distribution in each bin,   
%in a given interval $\Delta = 1$ along the $L_z$ axis.  
%In physical units each interval is simply a bin of width $\hbar$ 
%about one of the values $m_l \in \{ -l\hbar , (-l+1)\hbar , 
%\dots, l \hbar \}$. 
If this relative measure approaches zero in the classical limit then 
the quantum probabilty distribution converges  
to the corresponding classical one in that limit. 

\begin{figure}
\begin{center}
\input{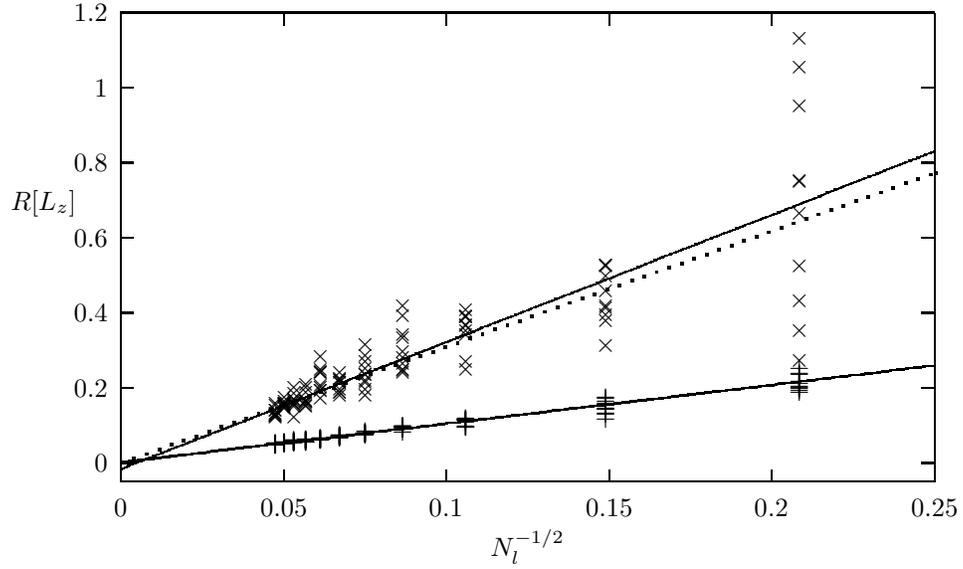}
\vspace{0.2in}
\caption{
Scaling of relative quantum-classical differences (\ref{eqn:rlz}  
in the equilibrum time-domain versus increasing system size. 
Scatter of crosses corresponds to time-steps  $191 \leq n \leq 200$,  
for a state launched in 
the chaotic zone of the mixed regime ($\gamma=1.215$). 
Scatter of plus signs corresponds to time-steps  $41 \leq n \leq 50$,  
for a state launched in the global chaos regime. 
Data sets in both of these regimes are consistent with the scaling law 
$R \simeq {N_l}^{-1/2}$, where $N_l =2l+1$.  
}
\label{Rvssqrtnl}
\end{center}
\end{figure}

%We are interested in characterizing the scaling of $R$ in 
%the equilibrium time-domain. 
%, that is, 
%once the distributions have relaxed close to their invariant 
%equilibrium distribution. 
In Fig.\ \ref{Rvssqrtnl} we consider typical equilibrium 
values of $R[L_z(n)]$ plotted against $1/\sqrt{N_l}$.  
We study the scaling using $N_l$ because it 
is equal to the dimension of the factor space ${\mathcal H}_l$ and 
it is also proportional to the subsystem size $N_l \simeq 2 |{\bf L}|$. 
We first consider a state launched in the global chaos regime 
($\gamma=2.835$, $r=1.1$), with initial condition 
$\theta(0) = (45^o,70^o,135^o,70^o)$. The scatter of plus 
signs for each $N_l = 2l+1$ value 
corresponds to time-steps $n$ such that 
$41 \leq n \leq 50$.  These time-step values are 
chosen because they occur well after the relaxation time 
$t_{rel} \simeq 6$. In this regime the data exhibits very 
little scatter. A least-squares 
fit to the curve $ R = A / \sqrt{N_l} + B $ yields a value for the 
intercept $B$ that is 
consistent with zero ($ B =0.001  \pm 0.001$) and a slope of order 
unity ($A = 1.032 \pm 0.02$). 
% with a reduced $\chi^2_\nu \simeq 1$. 
An intercept consistent with zero implies   
that quantum-classical differences vanish in the classical limit, {\it i.e}.\ 
$P_{L_z}(m_l) \rightarrow P_{L_z}^c(m_l)$ as $l \rightarrow \infty$. 
This result is especially remarkable since we have considered 
the differences that arise given classical measurements  
which resolve the observable $L_z$ with the rather extraordinary 
precision of $\hbar=1$.  

We next consider a state launched from the chaotic zone of 
the mixed regime ($\gamma=1.215$, $r=1.1$) 
with $\theta(0) = (20^o,40^o,160^o,130^o)$.
The scatter of crosses in  Fig.\ \ref{Rvssqrtnl} corresponds to 
time-steps 
$191 \leq n \leq 200$ $n$, again chosen well after the relaxation time 
$ t_{rel}$ for the range of quantum numbers considered. 
The scatter of quantum-classical differences at each $N_l$ value  
is much more significant in this regime in which the 
equilibrium distributions reflect a much more complex 
phase space structure. 
However, the relative differences exhibit, on average, a similar dependence 
on the quantum numbers as in the predominantly chaotic regime. 
In this regime a least-squares fit to the curve  
$ R = A / {N_l}^{1/2} + B $ yields a slope of order 
unity ($A = 3.39 \pm 0.15$) but a negative 
value for the intercept ($B=- 0.017 \pm 0.009$) within 
two-standard deviations of zero. 
A negative intercept is not physically meaningful 
(since $R$ is a positive definite 
quantity) and we assume it arises as a consequence 
of the statistical scatter in the data.  
Also plotted is the curve $R =  C/N_l^{1/2}$, 
with slope $C =3.09 \pm 0.04$ also determined from a least-squares fit. 
Both fits are good, with reduced $\chi^2$ values of order unity. 

%Although the initial state is a product of maximally localised 
%pure states, at later times each subsystem spin 
%must be described by a mixed state in its factor space. 
%This evolution from pure to mixed in the factor space  
%is a result of dynamically induced entanglement between 
%the quantum subsystems.  
%We have determined that the subsystems  
%remain entangled in the equilibrium time-domain 
%by confirming that the corresponding subsystem states 
%are nearly maximally mixed,  
%${\mathrm Tr}[ [ \rho^{(l)}(n) ]^2 ] \simeq (2s+1)^{-1} $ 
%for $n > t_{rel}$. 
%$\rho^{(l)}$ 
%we may expect that typical quantum fluctuations will 
%be reduced in magnitude relative to the fluctuations 
%occurring for pure states. Moreover, 

\begin{figure}
\begin{center}
\input{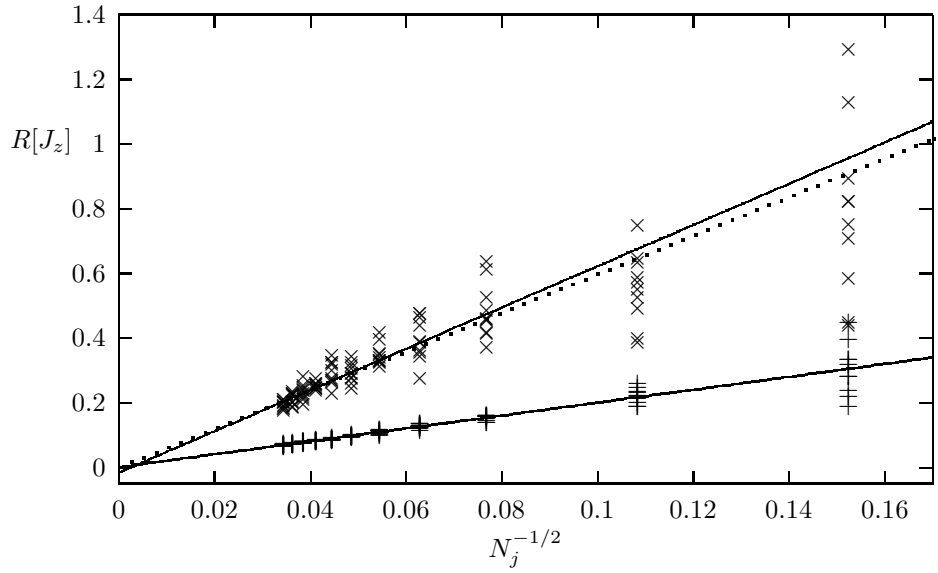}
\vspace{0.2in}
\caption{
Same as Fig.\ \ref{Rvssqrtnl} but using $R[J_z]$, as given by 
(\ref{eqn:rjz}). Data sets in both types of chaotic regime  
are consistent with the scaling law 
$R \simeq {N_j}^{-1/2}$, where $N_j = 2(l+s)+1$.  
}
\label{Rvssqrtnj}
\end{center}
\end{figure}

As we noted in the last section, the subsystem 
states do not remain pure, because of dynamically induced  
entanglement between the subsystems. 
Since the subsystem state (\ref{eqn:redl}) 
in the factor space ${\mathcal H}_l$ 
is not pure, but highly mixed in the equilibrium time-domain, 
it is possible that the scaling with $N_l$ 
that we observe is related to the purity-loss from 
this entanglement. 
% with a between subsystems. 
%with the ${\bf S}$ subsystem state.  
% since we are effectively 
%averaging over some mixture of pure state fluctuations. 
%rather than 
%a mixed state in the factor space
Consequently, 
it is useful to examine the scaling of the quantum-classical 
differences for the total spin $J_z$. 
The operator $J_z$ acts non-trivially in 
the full Hilbert space ${\mathcal H}$.
%; specifically, it has a non-degenerate 
%eigenvalue spectrum on ${\mathcal H}$. 
In this Hilbert space the system is described 
by a pure state vector at all times. 
In Fig.\ \ref{Rvssqrtnj} we consider the scaling of the ratio,   
\begin{equation}
\label{eqn:rjz}
R [J_z(n)] =  \frac{D [J_z(n)]}{{\overline P}_{J_z}}  
= N_j \; \sigma[J_z(n)],  
\end{equation}
where ${\overline P}_{J_z}  = [2(s+l)+1]^{-1}= N_j^{-1}$ is the 
average value of 
either distribution,    
versus the dimension $N_j$.  Here $N_j$ is the number of subspaces 
associated with distinct eigenvalues ($m_j$) of the quantum 
operator ($J_z$). In contrast with $N_l$, $N_j$ is not equal to 
the dimension of the corresponding Hilbert space, though it 
is a measure of the system size since $N_j \simeq 2 |{\bf J}|$. 
%observable $m_j$ of the spin component $J_z$. 
%Each eigenvalue is associated 
%with a subspaces that span 
%and the root-mean difference $D [J_z(n)]$ both 
%decrease as $l \rightarrow \infty$. 
The parameters and initial conditions shown  in  Fig.\ \ref{Rvssqrtnj} 
are the same as in Fig.\ \ref{Rvssqrtnl}. 
The same fit procedure as above, but applied to the 
function $R =  A/N_j^{1/2} + B$,  yields a  value for $B$
that is again consistent with zero ($B= 0.00038 \pm 0.0016$) and a 
positive slope of order unity ($A =  2.00 \pm 0.04$)
in the predominantly chaotic regime (scatter of plus signs). 
Thus the relative standard deviation for $J_z$ 
%$R [J_z(n)]$ 
also decreases as the square of the quantum numbers 
and fits to an intercept that is consistent with zero. This implies 
that the fluctuating quantum distributions 
approach the classical equilibrium, even for a few degree-of-freedom  
system, which is described at all times by a pure state.
In a chaotic state of the mixed regime (scatter of crosses), the 
fluctuations are larger, and the same fit 
procedure as above gives ($B = -0.016 \pm 0.012$, $A = 6.4 \pm 0.3$), 
where the negative value for $B$ lies within two standard deviations of 
zero and is presumed to result from the statistical 
scatter of the data.    
Also plotted is the equation $R =  C/N_j^{1/2}$, with $C =5.97 \pm 0.06$ 
determined from a least-squares fit. The fits to both equations 
are good, with reduced $\chi^2$ values of order unity.

%The situation should be contrasted with that 
%for the operator $L_z$, which can be expressed as the operator 
%$L_z \otimes {\bf 1}_s $ 
%acting on ${\mathcal H}$. In the subspace 
%${\mathcal H}_l$ on which $L_z$ acts non-trivially the system 
%can only be described by a mixed state, with corresponding density 
%operator $\rho^{(l)}$ given  by (\ref{eqn:redl}). 

\section{Discussion}

We have shown that, in classically chaotic regimes, initially 
localised quantum states relax 
to an equilibrium configuration that reflects the details of 
the classical phase-space structure. 
%Under these conditions, 
%the Shannon entropy grows approximately linearly in time. 
%This relaxation occurs much more slowly in chaotic zones of 
%mixed regimes. 
%We have consider these signatures at the level of 
%quantum and classical probability distributions for 
%classical observables. 
We find a remarkable degree of correspondence between the quantum and 
classical relaxation rates, even for small quantum numbers. 
Moreover, contrary 
to results obtained for the low-order moments \cite{Ball98,EB00a}, 
the degree of difference between  
the probability distributions is actually smaller for the chaotic 
states than the regular states. 

%Under chaotic dynamics both the quantum and classical 
%distributions relax towards the same equilibrium configuration 
%at an approximately exponential rate. 
The equilibrium quantum distributions exhibit small rapidly oscillating  
fluctuations about the coarse-grained classical equilibrium. 
As the measure of regular islands on the classical manifold 
approaches zero, the quantum and classical equilibrium configurations  
approach their microcanonical forms, and the quantum fluctuations 
about the classical equilibrum approach a non-vanishing minimum. 
% when the measure of regular islands approaches zero (producing ergodic 
%dynamics on the accessible manifold). 
This minimum arises because we consider total quantum states 
that are pure, whereas the microcanonical configuration 
is produced by an equal-weight mixture. 

For the distributions associated with the 
subsystem observable ${\bf L}$, 
the scale-independent standard deviation of these 
differences decreases as $N_l^{-1/2}$ 
where $N_l=2l+1 \simeq 2 |{\bf L}|$ 
is the dimension of the factor space, 
and becomes vanishingly small in the limit 
of large quantum numbers ({\it i.e}.\ large spins). 
These results suggest that 
correspondence with classical Liouville mechanics emerges 
in the classical limit for 
time-scales much longer than the Ehrenfest time.  
%closed system described initially by localised pure states 
%that are subject to unitary evolution. 

A great deal of recent work has emphasized that  
the loss of purity resulting from interactions 
with a quantum environment removes characteristic 
quantum effects and improves the degree of quantum-classical 
correspondence \cite{HB96,Zurek98a,Brumer99}. 
%It has been suggested that this process of {\em decoherence}  
%is {\em necessary} in order to see the emergence of classical properties 
%from quantum mechanics.  
While this is certainly the case 
for small quantum systems, it has been further argued 
that these {\em decoherence} effects 
must be taken into consideration to see the emergence of classical 
properties from quantum mechanics,  
even in the limit of large quantum numbers, 
if the classical motion is chaotic \cite{ZP95a,Zurek98b}. 

Since our model is comprised of interacting subsystems,  
initially separable pure states become entangled dynamically;  
the subsystem states (in each factor space) 
do not remain pure but become mixed. 
This entanglement process has 
an effect that is analogous to the process of decoherence. 
Hence one might suspect that 
the emergent classical behaviour that we have observed 
for the properties of the subsystem ${\bf L}$ 
may be strictly the result of a ``decoherence'' effect arising 
from entanglement with the other subsystem.
%It should be noted that such an effect would be essentially new 
%since $S$ is a few degree-of-freedom quantum object subject to 
%chaotic dynamics rather than a many-degree-of-freedom environment. 
To address this possibility, we have 
considered also the quantum-classical differences that arise 
in the probability distributions for a total system observable, $J_z$. 
In this case the quantum observables are 
projectors onto subspaces of the full Hilbert space, rather 
than merely a factor space.  The quantum state 
in this full Hilbert space is not subject to any entanglement 
or decoherence and remains pure throughout the unitary time-evolution.  
%As expected, the standard deviation of the differences 
%associated with the total spin  
%observables is larger than for the subsystem observables 
%(when comparing pure total system states and mixed subsystem states 
%with the same Hilbert space dimension), 
%and the loss of purity that arises from entanglement between 
%subsystems does have a decreasing effect on the characteristic 
%quantum fluctuations. More significantly, 
We have found that the scale-independent standard deviations for 
these quantum-classical 
differences decrease as $1/\sqrt{N_j}$, where $N_j=2 (s+l) +1$ 
is a measure of the system size,  $N_j \simeq 2 |{\bf J}|$.  
%Since we keep $l \simeq s$, it follows that 
%$N_j \simeq 2 \sqrt{N}$  
where $N = (2s+1)(2l+1)$ is the dimension of the Hilbert space.   
%bins of width $\hbar=1$ in the probability distribution.  
%along the $J_z$ axis.. 
%Since we always keep $l \simeq 1.1 s$, the Hilbert space dimension 
%is $\mathcal O(N^2)$. 
%It may be important to note that these total angular momentum 
%observables, unlike the subsystem observables, are obtained from 
%the sum (\ref{eqn:pjz}) and, consequently, are subject to some classical 
%averaging.  In spite of this caveat, our results suggest 
%, for the chaotic states of our model (\ref{eqn:ham}), 
The bin-wise quantum-classical differences 
become increasingly difficult 
to observe, in the limit of large quantum numbers, even for 
system observables that are isolated from the effects of decoherence.
%Of course the additional coarse-graining that arise given  
In this sense the process of decoherence 
%, though possibly sufficient,  
%associated with entanglement and the resultant loss of purity 
%the resulting impurity of the reduced state operator) 
is not {\em necessary} to produce quantum-classical  
correspondence in the classical limit. 
%at either early or late times in the limit of lage quantum numbers. 

%An interesting result we have not pursued is to study 
%how the characteristic fluctuations in the factor space 
%states (which are partially mixed from dynamically induced entanglement) 
%scale when the dimension of the other factor space is varied independently. 

%Decoherence has been  
%shown to have a significant decreasing effect 
%on the quantum-classical differences associated with a fixed value 
%of $\hbar$. 
%(With $\hbar$ fixed the dimension of the factor space is fixed but 
%the dimension of the full Hilbert space increases with the addition 
%of auxiliary quantum systems. This growth of the 
%Hilbert space dimension is a thermodynamic, 
%rather than classical, limit). 
%However, the correspondence results we obtained were not limited to 
%the quantum-classical differences associated with 
%the probability distributions for observables in one of the factor spaces. 

\section{Acknowledgements}

We thank the Natural Sciences and Engineering Research Council of Canada 
for financial support. 
J.\ E.\ would like to thank K.\ Kallio for stimulating discussions 
and the Centre for Experimental and Constructive Mathematics 
at Simon Fraser University for access to computational resources.

%\input{schnirel}

% now the references. delete or change fake bibitem. delete next three
%   lines and directly read in your .bbl file if you use bibtex.

\newpage
\vspace{0.5in} 

\newpage
\vspace{0.5in} 

\vspace{0.5in} 
\newpage

\newpage
\vspace{0.5in} 

\newpage
\vspace{0.5in} 

\newpage
\vspace{0.5in} 

\newpage
\vspace{0.5in} 

\newpage
\vspace{0.5in}

\newpage
\vspace{0.5in}

\newpage
\vspace{0.5in}

\newpage
\vspace{0.5in}

%\include{extrafigs}

% \begin{table}
% \caption{}
% \label{}
% \end{tabular}
% \end{table}

\end{document}